\documentclass[nofootinbib,prl,floatfix,preprintnumbers,amsmath,amssymb,groupedaddress,superscriptaddress,twocolumn]{revtex4}
\usepackage{psfrag,graphicx}

\usepackage{dsfont}
\usepackage{epsfig}

\def\be{\begin{equation}}
\def\ee{\end{equation}}
\def\bea{\begin{eqnarray}}
\def\eea{\end{eqnarray}}

\begin{document}
\title{The   Holographic Superconductor Vortex}
\author{Marc Montull}
\email{mmontull@ifae.es}
\affiliation{Departament  de F\'isica and IFAE, Universitat Aut{\`o}noma de Barcelona,
08193 Bellaterra, Barcelona}

\author{Alex Pomarol}
\email{alex.pomarol@uab.cat}
\affiliation{Departament  de F\'isica and IFAE, Universitat Aut{\`o}noma de Barcelona,
08193 Bellaterra, Barcelona}

\author{Pedro J. Silva}
\email{psilva@ifae.es}
\affiliation{Departament  de F\'isica and IFAE, Universitat Aut{\`o}noma de Barcelona,
08193 Bellaterra, Barcelona}
\affiliation{Institut de Ci\`encies de l'Espai (CSIC) and Institut
d'Estudis Espacials de Catalunya (IEEC/CSIC),
Universitat Aut{\`o}noma de Barcelona,
08193 Bellaterra, Barcelona}

\begin{abstract}
A  gravity dual of a superconductor at finite temperature  has been recently proposed.
We present the vortex configuration of this model  and study its properties.
In particular,  we calculate  the   free energy
as a function of an external magnetic field,  the
magnetization and the superconducting density.
We also find the two critical magnetic fields that define the region in which
 the vortex configurations are  energetically favorable.
\end{abstract}
\maketitle

\section{Introduction}

The Gauge/Gravity duality, that relates strongly interacting gauge theories to theories of gravity in higher dimensions, has  opened  a new window to study many different strongly interacting systems.
The applicability of this  approach is very vast ranging from
particle physics to plasma and  nuclear physics.
In Ref.~\cite{Hartnoll:2008vx} a model for a dual description of a superconductor was proposed.
The model showed to have a critical temperature $T_c$ under which the system
goes into a superconducting phase.
The properties of this phase have been thoroughly   studied \cite{Hartnoll:2009sz}, showing a resemblance with those of a  Type II superconductor.
In spite of this,   Abrikosov vortices, known to happen in
Type II superconductors, have not yet been obtained.
The purpose of this letter is to show that  in this type of gravity duals
  vortex solutions indeed exist and can be energetically favorable in the presence
 of  external magnetic fields.
  Due to the nonlinear nature of these configurations, we  will have to rely on numerical
methods. Among other physical properties,   we will calculate the free energy
 and the  range of the magnetic field $B_{c\, 1}\leq B \leq B_{c\, 2}$ at which the  superconductor is   at the  intermediate phase (Shubnikov phase)  characterized  by vortex configurations.
Further aspects  of these solutions will be presented elsewhere.

\section{The Model}
The physical system to study is a conformal strongly coupled superconductor in 3D at finite temperature and charge density.
Its gravitational dual theory \cite{Hartnoll:2008vx}
is an asymptotically AdS-Schwarzschild space-time in 4D. The gravitational degrees of freedom are coupled to an U(1) gauge field $A_\mu$ and a complex scalar $\Psi$. The action that summarizes the above model is given by
\bea
&&S=\int d^4x\, \sqrt{-G}\left\{\hbox{${1\over16\pi G_N}$}\left(R+\Lambda\right)\,-\frac{1}{g^2}\mathcal{L}\right\}\, ,\nonumber\\
&&\hbox{with }\mathcal{L}={1\over4}F^2+\frac{1}{L^2}|D_\mu\Psi|^2+\frac{m^2}{L^4}|\Psi|^2\, .
\label{action}
\eea
$G_N$ is the 4D gravitational Newton constant, the cosmological constant $\Lambda$ defines the asymptotic AdS radius $L$ via the relation $\Lambda=-{3/L^2}$ and
$D_\mu=\partial_\mu-iA_\mu$. We use the convention where the metric $G$ has signature $(-,+,+,+)$, with coordinates $(t,z,r,\phi)$ where $t$ is time, $z$ is the holographic direction such that the AdS-boundary  occurs at $z=0$, and  $(r,\phi)$ are polar coordinates parameterizing the remaining 2D plane. For the scalar mass $m^2$ we will focus on two possible values: $m^2=-2,0$.
Other values are expected to give similar behaviors \cite{Horowitz:2008bn}.

We will work in the so-called probe approximation, where the gravity sector is effectively decoupled from the matter sector and therefore, there is no back-reaction on the background metric due to $\mathcal{L}$. This regime is achieved in the limit of large $g$, when compared to the gravitational strength. In this limit we can, without loss of generality, fix $g=1$.
   In our conventions, the background AdS-Schwarzschild Black hole (BH) metric is given by
\begin{equation}
ds^2=\frac{L^2}{z^2}\left(-f(z)dt^2+dr^2+r^2d\phi^2\right)+\frac{L^2}{z^2f(z)}dz^2\, ,
\end{equation}
where $f(z)=1-(z/z_h)^3$.

As we are considering the theory at finite temperature, we have to take the Euclidian regime with compact time $it\in [0,1/T]$  where $T=3/(4\pi z_h)$. Therefore, the holographic coordinate runs from the AdS-boundary at $z=0$ to the BH horizon at $z=z_h$. Notice that we work with a planar BH with energy per unit area $\varepsilon=L^2/(8\pi G_N z_h^3)$. Then, the AdS/CFT duality tells us that the above are precisely the temperature and energy density of the dual superconductor.

The gauge field has the usual AdS-boundary behavior
\be
A_\nu \rightarrow a_\nu+J_\nu z\, ,
\ee
where $a_\nu=(\mu,a_i)$ corresponds to  the potentials on the dual CFT, while $J_\nu=(-\rho,J_i)$ plays the role of the conjugated currents.
We will consider the case in which  the charge density $\rho$ is fixed constant.
The other potentials $a_i$ are related to turning on either electromagnetic fields  or sample velocities in the dual CFT, depending on the interpretation we give to the AdS/CFT duality. The first interpretation is what we will use in this article, while the second one is relevant for superfluids
\footnote{In fact, the vortex solution we present in this article can
be identified with vortex configurations in a superfluid, once the
appropriated reinterpretations are made.}
 \cite{Herzog:2008he}.
Similarly, the scalar field  has the following AdS-boundary behavior
\be |\Psi| \rightarrow az^{3-\Delta}+bz^{\Delta}\, ,
\label{psiads}
\ee
where $\Delta=2,3$  (for $m^2=-2,0$)
corresponds to the dimension of the dual operator ${\cal O}_\Delta$ responsible for the U(1) breaking, and
$b$ determines the  vacuum expectation value of this operator.
The value of $a$  corresponds to an explicit breaking of the U(1) symmetry and  will then be  turned to zero  \footnote{For the case $m^2=-2$ there is the possibility to have $b=0$ and
$a\not=0$ corresponding to have a dual CFT operator of dimension one \cite{Hartnoll:2009sz}.}.
Having fixed $m^2$, the only parameters of the model
are the scales $T$ and $\sqrt{\rho}$.

It has been reported in Ref.~\cite{Hartnoll:2008vx,Horowitz:2008bn}
 that  for $\rho\not=0$ the system undergoes a phase transition
at
\bea
T_c&\simeq& 0.12\sqrt{\rho}\  \ \ {\rm for}\ \ \ m^2=-2\, ,\nonumber\\
T_c&\simeq& 0.09\sqrt{\rho}\  \ \ {\rm for}\ \ \ m^2=0\, ,
\eea
 where the  two phases are related to  a charged BH and a charged BH with a non-trivial scalar hair.
 At $T<T_c$, the system is at
the hairy phase   corresponding   to a superconducting phase.
In Refs.~\cite{Hartnoll:2008kx,Albash:2008eh}  the model was also studied
 in the presence of an external magnetic field $B$  using a dyonic BH with a probe scalar field. The result was a bounded superconducting region or drop, that squeezes to zero size  as we increase $B$. The above suggested that we are dealing with a Type II superconductor.
 If this is the case, Abrikosov vortex configurations should be present in this model.

 We stress that, as is usual in this approach,
  we are  treating the electromagnetic field
of the  3D dual theory  as a nondynamical  background.
This corresponds to take the 3D electric charge  $e\rightarrow 0$, while keeping constant $B$ and $\rho$.

\section{The vortex solution}

We use the Ansatz given by
\begin{equation}
\Psi=\psi(r,z)\, e^{in\phi}\ ,\ \ A_0=A_0(r,z)\ , \ \  A_\phi=A_\phi(r,z)\, ,
\end{equation}
with all other fields set to zero. This Ansatz preserves global U(1) transformations when combined with a rotation in the 2D plane. The fields $A_r,A_z$ can be consistently set to zero since our Ansatz fulfills
$\partial_r Arg[\Psi]=\partial_z Arg[\Psi]=0$. The winding number $n\in Z$ determines different topological  solutions.
With the above  Ansatz we obtain  from Eq.~(\ref{action}) the following equations of motion:
\begin{eqnarray}
&&
z^2\partial_z \left(\frac{f}{z^2} \, \partial_z \psi \right)+\frac{1}{r} \partial_r\left(r\partial_r \psi \right) \nonumber \\
&&
\hspace{2,2cm}+\left(\frac{A^2_0}{f}-\frac{(A_\phi-n)^2}{r^2}-{m^2\over z^2}\right) \psi=0\, ,\nonumber \\
&&
\partial_z \left( f\partial_z A_\phi \right) + r \, \partial_r \left( \frac{1}{r}
 \partial_r A_\phi \right) - \frac{2\psi^2 }{z^2} (A_\phi-n) = 0\, ,\nonumber\\
&&
f\partial^2_z A_0 + \frac{1}{r} \partial_r \left( r \partial_r A_0 \right ) - \frac{2\psi^2}{z^2} A_0= 0\, .
 \label{pdes}
\end{eqnarray}
In order to  describe a dual  superconductor at fixed $\rho$  in the presence of an external magnetic field  $B$,  the AdS/CFT correspondence tells us that we must impose the
 AdS-boundary conditions
\begin{equation}
\psi|_{z=0}=0\ ,\ \  \partial_z A_0|_{z=0}=-\rho\ , \ \  A_\phi|_{z=0}=\frac{1}{2}r^2B\, ,
\label{adsbc}
\end{equation}
for the case $m^2=0$, while  for $m^2=-2$ the first condition must be  $\partial_z\psi|_{z=0}=0$ (this is equivalent to set $a=0$ in Eq.~(\ref{psiads})).
At the horizon $z=z_h$ we require the field configurations to be regular; in particular we set $A_0|_{z=z_h}=0$ as usual, to have a well-defined Euclidean continuation. Similar reasoning at $r=0$ implies  that  for $n\not=0$
\begin{equation} \psi|_{r=0}=0\ ,\ \  \partial_r A_0|_{r=0}=0\ , \ \  A_\phi|_{r=0}=0\, ,
\end{equation}
while  for $n=0$,   $\partial_r \psi|_{r=0}=0$.
We will be considering a 3D superconductor of radius $R$
that we will take to be much bigger than the vortex radius.
This is implemented by setting a nonzero $\rho$ extending  from $r=0$ to $r=R$.

The 2D system of the three partial differential equations   of Eq.~(\ref{pdes})
is  nonlinear, and therefore requires to be  solved numerically.
For this purpose we have used the COMSOL 3.4 package \cite{comsol}.
In our numerical studies we have chosen
\be
R=\frac{50}{\sqrt{\rho}}\ ,\ \ \     T= 0.065\sqrt{\rho}\, .
\label{input}
\ee
This corresponds to
\be
\frac{T}{T_c} \simeq 0.74\ (0.54)\, ,
\ee
for the case of $m^2=0\ (-2)$.

In Fig.~1 we show  the order parameter
$\langle {\cal O}_\Delta\rangle=\frac{1}{\Delta}z^{1-\Delta}\partial_z\psi|_{z=0}$ of the dual superconductor.
 We can see that this goes to zero at the origin where the vortex is placed.
For the value of the magnetic field, we have chosen
\be
B_n= \frac{2n}{R^2}\, ,
\label{bn}
\ee
corresponding to the
 value at which the  magnetic flux crossing a surface of constant $z$,
 $\Phi=\int d\phi \int^R_0 r dr B$, equals $2\pi n$.
This is the quantized flux going through the $n$-vortex of the dual superconductor.

\begin{figure}[ht]
  \centering
  \includegraphics[width=7.5cm]{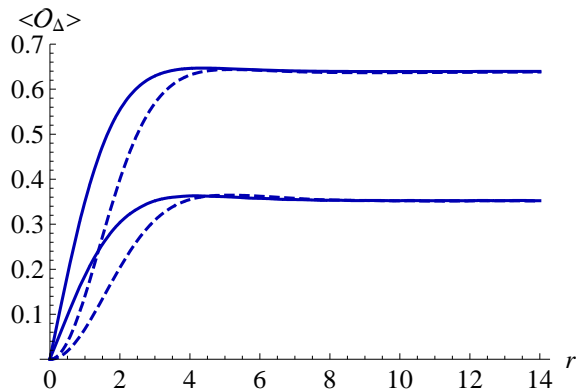}
  \caption{\textit{Order parameter $\langle {\cal O}_\Delta \rangle$
   for the $n=1$ (solid) and
$n=2$  (dashed) vortex configuration. The lower (upper) curves correspond to the case
   $m^2=0\ (-2)$.  Presented in units of $\sqrt{\rho}=1$.}}
\end{figure}

\section{Free energy, magnetization  and Critical Magnetic fields}

We are interested to determine the free energy of the superconductor configurations with $n=0,1,2$ to know which one is  energetically favorable as we vary $B$.
By the AdS/CFT, the free energy $F$ of the superconductor  is given by
\begin{equation}
\frac{F[T,B,\rho]}{T}= S_E+\frac{\pi}{T} \left. \int^R_0  {dr}{r} A_0\partial_z A_0 \right|_{z=0}\, ,
\label{freee}
\end{equation}
where the right-hand side is evaluated on-shell in the 4D theory with the boundary conditions
given in Eq.~(\ref{adsbc}).
The second  term of Eq.~(\ref{freee}) has  been  added to guarantee the variational principle
when working   at fixed  $\partial_z A_0$ on the AdS-boundary.
Since, as we will see, the phase transition to vortex configurations occurs at small values of $B$, we can treat the magnetic field as a small perturbation and separate the solution as
\begin{equation}
\psi\rightarrow \psi+\delta \psi\ ,\  \ A_0\rightarrow A_0+\delta A_0\ ,\ \
A_\phi\rightarrow A_\phi+\delta A_\phi\, ,
\end{equation}
where  the unperturbed  solution $(\psi,A_0,A_\phi)$ corresponds to that at zero external magnetic field, {\it i.e.},
$A_\phi|_{z=0}=0$,   while the perturbation  $(\delta \psi, \delta A_0, \delta A_\phi)$
must fulfill
\be
 \delta A_\phi|_{z=0}=\frac{1}{2}r^2B\ ,\ \partial_z \delta A_0|_{z=0}=0\ ,\
\delta \psi|_{z=0}=0 \, ,
\ee
for $m^2=0$ and  $\partial_z\delta \psi|_{z=0}=0$ for  $m^2=-2$.
By  integrating by parts the free energy of the $n$-vortex configuration can be written,
up to  $B^2$ terms, as

\begin{equation}
F_n(B)\simeq F_n(0)-\alpha_n B+\frac{1}{2}\beta_n B^2\, ,
\label{freeapprox}
\end{equation}
where we have defined
\begin{eqnarray}
F_n(0)&=&
 2\pi \int^R_0 dr \int^{z_h}_0 dz \, \frac{r}{z^2} \left( \frac{A_0^2}{f}-\frac{A_\phi
 (A_\phi-n)}{r^2} \right) \psi^2 \nonumber\\
&-&\pi \left. \int^R_0 dr r A_0 \, \partial_z A_0 \right |_{z=0}\, ,\nonumber\\
\alpha_n&=&  \frac{2\pi}{B} \left. \int^R_0 \frac{dr }{r}  \delta A_\phi \partial_z A_{\phi}  \right |_{z=0} \, ,\nonumber\\
\beta_n&=&- \frac{2\pi}{B^2} \left. \int^R_0 \frac{dr }{r}  \delta A_\phi \partial_z \delta A_{\phi}  \right |_{z=0}\, .
\label{fab}
\end{eqnarray}
Notice that the positive-defined quantities $\alpha_n$ and $\beta_n$  do not depend on  $B$,
 since $ \delta A_\phi \propto \delta A_\phi|_{z=0}\propto B$.
Eq.~(\ref{freeapprox}) has a simple interpretation in terms of the magnetization $M$ of the superconductor.
Using  $M=-\partial F/\partial B$, we can write
\begin{equation}
F_n(B)=F_n(0)-\int^B_0 M_n dB\, ,
\label{newform}
\end{equation}
where the magnetization of the $n$-vortex configuration $M_n$ in the $z$-component is given by
\begin{equation}
M_n=\frac{1}{2}\int d\phi\, dr\, r (\vec r\times \vec J)_z=\pi\int dr\, r  J_\phi\, .
\label{mn}
\end{equation}
From the AdS/CFT dictionary, we have that
\begin{equation}
\langle J_\phi\rangle=-\left.\frac{\delta F}{\delta A^\phi|_{z=0}}= \partial_z A_\phi+\partial_z \delta A_\phi\right|_{z=0}\, ,
\end{equation}
that together with Eq.~(\ref{mn}) leads to our final expression for the magnetization
\begin{equation}
M_n=\alpha_n-\beta_n B\, .
\end{equation}
Using this  expression  into Eq.~(\ref{newform}), we recover the free energy  of Eq.~(\ref{freeapprox}).

For the free energy at $B=0$ we obtain
\begin{equation}
F_n(0)\simeq F_0(0)+0.9(1.5)  n^2\ln [R\rho^{1/2}]\sqrt{\rho}+c_n\, ,
\end{equation}
where $c_0=0$, $c_1\simeq 1.2(3.7)\sqrt{\rho}$, $c_2\simeq 0.3(4)\sqrt{\rho}$ and
\begin{equation}
F_0(0)\simeq  5 (4) R^2\rho\sqrt{\rho}\, ,
\end{equation}
for the case $m^2=0(-2)$.
This shows that, as expected, the vortex configurations have for $B=0$ a larger energy than the $n=0$ solution.
Note that  $F_0(0)$  grows  with the volume of the superconductor ($\propto R^2$), although not the difference $F_{1,2}(0)-F_0(0)$  that is only logarithmically sensitive to $R$ for $R\rightarrow \infty$, as expected for 3D vortices in the absence of  electromagnetic fields.
For the magnetization  we find
\begin{equation}
\alpha_n\simeq 0.4(0.7)\, n R^2\sqrt{\rho}\  , \ \ \ \beta_n\simeq 0.05(0.09) R^4\sqrt{\rho}\, .
\end{equation}
From Eq.~(\ref{freeapprox}) it is clear that there is a critical value for $B$ at which
the difference between the free energies $F_1(B)-F_0(B)$ is zero. This value is usually referred as $B_{c\, 1}$ and marks the beginning of the mixed phase where the magnetic field starts to penetrate the superconductor.
For the case of $m^2=0$ we have
\begin{equation}
{F_1(B)-F_0(B)\over \sqrt{\rho}}\simeq 0.9\ln [R\rho^{1/2}]+1.2-0.8\frac{B}{B_{1}}\, ,
\end{equation}
that for $R=50/\sqrt{\rho}$ equals to zero at
\begin{equation}
B_{c\, 1}\simeq 6 B_{1}\, ,
\label{bc1}
\end{equation}
where $B_1$ is defined in Eq.~(\ref{bn}).
For $m^2=-2$ we get  similar values, $B_{c\, 1}\simeq 7 B_1$.
At higher magnetic field values than $B_{c\, 1}$ the vortex configuration is preferred.
Notice that for $R\rightarrow\infty$, we have $B_1\rightarrow 1/R^2$ and therefore
$B_{c\, 1}\rightarrow 0$, indicating that the non-vortex solution
is never favorable at any $B\not=0$.

For the  configuration with $n=2$, we find that its free energy  is less than  that for $n=0,1$
if $B\gtrsim 10(14)  B_1$ for  $m^2=0(-2)$.
At this  high  magnetic field, however, we expect that the free energy  of a solution with two $n=1$ vortices will be  energetically more favorable, as it happens in Type II superconductors. Indeed, for two vortices sufficiently separated we expect
\bea
F(B)&\simeq& F_0(0)+2[(F_1(0)-F_0(0))-\alpha_1B]\nonumber \\
&+& E_{\rm int}+\frac{1}{2}\beta_1 B^2\, ,
\label{2v}
\eea
where $E_{\rm int}$ is the interaction energy between the two vortices.
Therefore  the difference between the free energy of two $n=1$ vortices  and  one $n=2$ vortex goes as  $\Delta F\simeq E_{\rm int}-1.8(3)\ln [R\rho^{1/2}]\sqrt{\rho}$ for  $m^2=0(-2)$.
As a consequence a configuration with two $n=1$ vortices will be preferred  for $E_{\rm int}<1.8(3)\ln [R\rho^{1/2}]\sqrt{\rho}$ that is expected for a  large superconductor.

On the other hand, as $B$ increases from $B_{c\, 1}$, a configuration with more and more vortices is expected to be favorable, until we reach a certain critical value $B_{c\, 2}$ at which there is another phase transition;   for $B>B_{c\, 2}$ the normal phase is preferred.
We estimate this value by the magnetic field  at which the superconducting region
of the $n=0,1$ configurations shrink to zero size.
We find $B_{c\, 2}\simeq 3(5)\rho$ for  $m^2=0(-2)$.

In Fig.~2 we plot the values of the free energy as a function of $B$
for the configurations $n=0,1,2$ from the exact numerical solutions.
We can see that the critical magnetic values at which the lines cross
  are similar to the approximate ones given above.

\begin{figure}[ht]
  \centering
  \includegraphics[width=7.5cm]{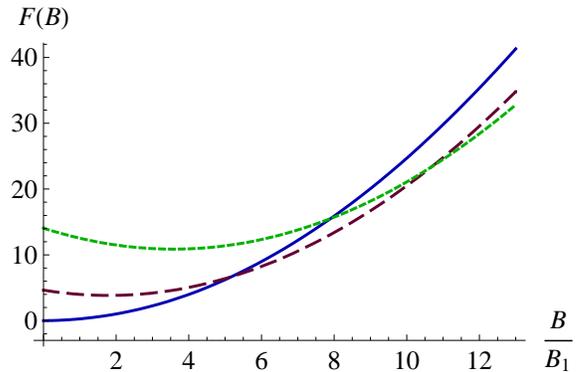}
  \caption{\textit{Free energy  for the $m^2=0$ case
  as a function of the external magnetic field
   for the $n=0$ (solid), $n=1$ (dashed)
  and $n=2$ (dotted)  vortex configuration. Presented in units of $\sqrt{\rho}=1$.}}
\end{figure}

Finally, we calculate the ``superconducting  density" $n_s(r)$ defined as
\begin{equation}
n_s(r)=\langle J_\phi J^\phi\rangle=\frac{\delta F}{\delta  A^{2}_{\phi}|_{z=0}}=-\left. \frac{\partial_z \delta A_\phi}{\delta  A_\phi}\right|_{z=0}\, ,
\end{equation}
where in the last equality we have used Eq.~(\ref{freeapprox}).
In Fig.~3 we show  $n_s(r)$ for the different configurations.
We notice that the vortex configuration fullfills
$\langle J_\phi\rangle= -n_s(r)(\delta A_\phi|_{z=0}-n)$,  as expected from
a spontaneously broken U(1) symmetry.
For a  non-vortex configuration  the superconducting density is constant
$n_s(r)\simeq 0.28(0.48) \sqrt{\rho}$ for $m^2=0(-2)$. This determines the  penetration length
$\lambda={1}/{(e \sqrt{n_s})}$  where $e$ is the electric charge of the dual superconductor.

\begin{figure}[ht]
  \centering
  \includegraphics[width=7.5cm]{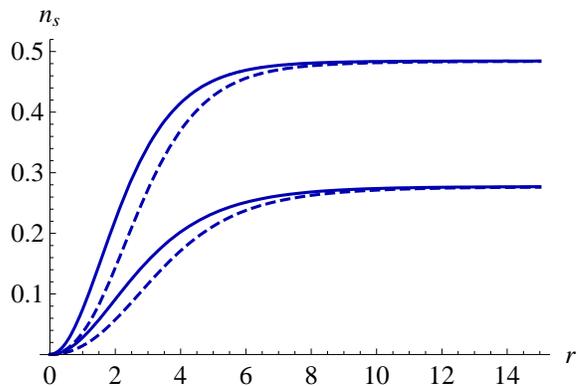}
  \caption{\textit{Superconducting density  $n_s(r)$ for the $n=1$ (solid) and
$n=2$  (dashed) vortex configuration. The lower (upper) curves correspond to the case
   $m^2=0\ (-2)$. Presented in units of $\sqrt{\rho}=1$.}}
\end{figure}

\noindent {\bf Note Added:}  While finishing  this paper,  we learned of
Ref.~\cite{Albash:2009ix}  which has also studied the vortex solution in holographic superconductors.

\noindent {\bf Acknowledgments:}  We would like to thank Alberto Salvio, Massimo  Mannarelli
and Alvar Sanchez for discussions.
The  work  of AP  was  partly supported   by the
Research Projects CICYT-FEDER-FPA2005-02211,
SGR2005-00916,  UniverseNet (MRTN-CT-2006-035863),
and  AP2006-03102.
The  work  of PJS  was  partly supported by the Research Projects CICYT-FEDER-FPA2005-02211 and FIS2006-02842, CSIC under
the I3P program.



\end{document}